\documentclass[12pt]{article}
\usepackage{amsfonts}
\newcommand{\p}[1]{(\ref{#1})}
\newcommand{\cp}{\mbox{$\cal P$}}
\newcommand{\e}{\eta}
\newcommand{\be}{\begin{equation}}
\newcommand{\bea}{\begin{eqnarray}}
\newcommand{\ee}{\end{equation}}
\newcommand{\eea}{\end{eqnarray}}
\newcommand{\la}[1]{\langle S_{#1}| }
\newcommand{\ra}[1]{|S_{#1}\rangle }

\def\theequation{\arabic{section}.\arabic{equation}}
\begin{document}
\setcounter{page}0
\renewcommand{\thefootnote}{\fnsymbol{footnote}}
\thispagestyle{empty}
{\hfill  Preprint JINR E2-2001-4}

\vspace{1.5cm}

\begin{center}
{\large\bf
On the mixed symmetry irreducible representations of the Poincare
 group in the BRST approach
}\vspace{0.5cm} \\

{\v{C}estm\'{\i}r Burd\'{\i}k}\footnote{E-mail: burdik@siduri.fjfi.cvut.cz}
\vspace{0.5cm} \\

{\it Department of Mathematics,
Czech Technical University,\\
Trojanova 13, 120 00 Prague 2}\vspace{0.5cm} \\

A. Pashnev${}^{a}$\footnote{E-mail: pashnev@thsun1.jinr.dubna.su}
and
M. Tsulaia${}^{a,b}$\footnote
{E-mail: tsulaia@thsun1.jinr.ru}\vspace{.1cm}\\

\vspace{0.5cm}
${}^a${\it Bogoliubov Laboratory of Theoretical Physics, JINR} \\
{\it Dubna, 141980, Russia}\\
~\\
${}^b${\it The Andronikashvili Institute of Physics, Georgian Academy
of Sciences,}\\
{\it Tbilisi, 380077, Georgia} \vspace{1.5cm}\\

{\bf Abstract}
\end{center}
\vspace{1cm}

The lagrangian description of irreducible massless representations
of the Poincare group with the corresponding
Young tableaux having two rows
along with some explicit examples including the notoph and
Weyl tensor
 is given. For this purpose
is used the method of the BRST constructions adopted to the
systems of second class constraints by the construction of
an auxiliary representations of the algebras of constraints
in terms of Verma modules.

\newpage\renewcommand{\thefootnote}{\arabic{footnote}}
\setcounter{footnote}0\setcounter{equation}0
\section{Introduction}

It is well known that the particles with the value of spin
more than two arise naturally when quantizing such classical objects as
the relativistic oscillator, string or discrete string. The challenging
problem  for these kinds of theories is to construct the lagrangian
description both for free and for interacting particles with
the higher spins.

Recent developments in this activity have revealed, that the particles
with the higher spins can propagate through the background
having the constant curvature, in particular through the AdS space
(see \cite{MV} and the references therein),
 as well as interact with the constant electromagnetic
field or symmetrical Einstein spaces  \cite{KL}.
The utilization of the technique of the Supersymmetric Quantum
Mechanics leads also to the description of the particle with spin 2
on the background of the constant curvature \cite{KY},
and on the background being real ``Kahler -- like" manifold
\cite{DP}.
In all these approaches corresponding
lagrangians possess some gauge   invariance
in order to remove the states
with the negative norm -- the ghosts -- from
the physical spectrum.

The same problem, i.e., the inclusion into the theory
of necessary gauge invariance is present also in the description
of free particles which belong to  irreducible representations
of the Poincare group.
 The possible way out of this problem is
the method of the BRST constructions which naturally leads to the
desired gauge invariant hermitian lagrangians \cite{OS}.
However, the construction of
the corresponding nilpotent BRST charges for the system of constraints
describing either reducible massive \cite{PT1} or irreducible massless
\cite{PT2}
representations of the Poincare group with the Young tableaux
having one row is not straightforward due to the presence of
the second class constraints.
This problem can be solved after the introduction into the theory of
the additional bosonic oscillator and the construction of
the auxiliary representations for the second class constraints in terms
 of this  oscillator.  Then after the partial gauge fixing
\cite{PT2} the
BRST invariant lagrangian leads to the one constructed by Fronsdal
 \cite{F} for the irreducible massless higher spins.

~The ~construction ~of ~lagrangians ~describing ~an ~arbitrary
~representations ~of ~the ~Poincare ~group ~is
~complicated ~due ~to the necessity of
the construction of the auxiliary representations of the Lie
algebras having the rank more than one in the framework of the BRST
approach. The general method for such constructions
where some generators from the Cartan subalgebra are excluded from the
total system of constraints forming an arbitrary Lie algebra was given
in \cite{BPT}. Below we use this method for the derivation
of the lagrangians for the massless particles belonging to the
irreducible representations of the Poincare algebra with the
corresponding Young tableaux having two rows.
 The complete generalization
of this procedure to the arbitrary irreducible representation of the Poincare group
will be given elsewhere.
Let us note that the approach used  is different from the
one given in \cite{LA} and leads to the different form of the
final lagrangian which describes only irreducible representations,
corresponding to  Young tableaux with two  rows.

The paper is organized as  follows.

In Sec.2 we quote the main results of \cite{BPT} needed for our study.

In Sec.3 we explicitly derive the lagrangians describing
the irreducible representations of the Poincare group
with corresponding Young tableaux having two rows.

In Sec.4 we present some explicit examples of our constriction.

Sec.5  contains our conclusions and discussion of open problems.
Some complicated expressions and formulae are carried out to Appendix.

\setcounter{equation}0\section{The BRST construction}
The fields we are going to consider  correspond to the following
Young tableaux

\begin{equation}\label{young}
\begin{picture}(65,20)
\unitlength=1mm
\put(-20,10){\line(1,0){65.2}}
\put(-20,5){\line(1,0){65.2}}
\put(-20,0){\line(1,0){55}}
\put(-20,0){\line(0,1){10}}
\put(-15,0){\line(0,1){10}}
\put(-10,0){\line(0,1){10}}
\put(5,0){\line(0,1){10}}
\put(10.2,0){\line(0,1){10}}
\put(15,0){\line(0,1){10}}
\put(20,0){\line(0,1){10}}
\put(25,0){\line(0,1){10}}
\put(30,0){\line(0,1){10}}
\put(35.2,0){\line(0,1){10}}
\put(40,5){\line(0,1){5}}
\put(45.2,5){\line(0,1){5}}
\put(-19.5,7){$\mu_1$}
\put(-14.5,7){$\mu_2$}
\put(40,7){$\mu_{n_1}$}
\put(-19.5,2){$\nu_1$}
\put(-14.5,2){$\nu_2$}
\put(30,2){$\nu_{n_2}$}
\multiput(-7.5,7.5)(5,0){10}{\circle*{.5}}
\multiput(-7.5,2.5)(5,0){8}{\circle*{.5}}
\end{picture}
\end{equation}
and are described by the field
$\Phi_{\mu_1\mu_2\cdots\mu_{n_1}, \nu_1\nu_2\cdots\nu_{n_2}}(x)$
which is the
  $n_1+n_2$ ($n_1 \geq n_2$) rank tensor field
symmetrical with respect to the
permutations of each type indices.
The correspondence with a given Young tableaux implies, that
after the symmetrization of all indices of the first raw with one index
of the second raw the basic field vanishes, i.e.,
\begin{equation}
\label{sym}
\Phi_{\bf \{\mu_1\mu_2\cdots\mu_{n_1}, \nu_1\}\nu_2\cdots\nu_{n_2}}
(x)=0.
\end{equation}
Further, all traces of the basic field must vanish:
\begin{eqnarray} \label{trace3}
\label{trace1}
&&\Phi_{{\mathbf{\rho\rho}}\mu_3\cdots\mu_{n_1}, \nu_1\nu_2\cdots\nu_{n_2}}(x)
=0, \nonumber\\
\label{trace2}
&&\Phi_{{\bf \rho}\mu_2\mu_3\cdots\mu_{n_1},{\bf \rho}
\nu_2\cdots\nu_{n_2}}
(x)=0,   \\  \nonumber
&&\Phi_{\mu_1\mu_2\cdots\mu_{n_1},{\bf \rho\rho}\nu_3\cdots\nu_{n_2}}
(x)=0,
\end{eqnarray}

In addition, this field is subject to the following
system of equations, namely the mass shell and
transversality conditions
for each type of indices. In the massless
case we have
\begin{eqnarray}\label{mass}
&&p_{\rho}^2
\Phi_{\mu_1\mu_2\cdots\mu_{n_1}, \nu_1\nu_2\cdots\nu_{n_2}}(x)
=0,\\
\label{trans1}
&&p_{\bf \rho}
\Phi_{{\bf \rho}\mu_2\cdots\mu_{n_1}, \nu_1\nu_2\cdots\nu_{n_2}}(x)
=0,\\
\label{trans2}
&&p_{\bf \rho}
\Phi_{\mu_1\mu_2\cdots\mu_{n_1}, {\bf \rho}\nu_2\cdots\nu_{n_2}}(x)
=0.
\end{eqnarray}

To describe all irreducible representations of the Poincare group
simultaneously it is convenient to
~introduce ~an ~auxiliary ~Fock ~space ~generated ~by
~the ~creation ~and ~annihilation
~operators $a^{i+}_\mu,a^j_\mu$
with Lorentz index $\mu =0,1,2,...,D-1$ and additional internal
index $i=1,2$. These operators satisfy the following
commutation relations
\be
\left[ a^i_\mu,a^{j+}_\nu \right] =-g_{\mu \nu}\delta^{ij},\;\quad
g_{\mu \nu}=diag(1,-1,-1,...,-1),
\ee
where $\delta^{ij}$ is usual Cronecker symbol.

The general state of the Fock space
depends on the space-time coordinates $x_\mu$
\be \label{Fockvector}
|\Phi\rangle =\sum
\Phi_{\mu_1\mu_2\cdots\mu_{n_1}, \nu_1\nu_2\cdots\nu_{n_2}}
(x)
a^{1+}_{\mu_1}a^{1+}_{\mu_2}\cdots a^{1+}_{\mu_{n_1}}
a^{2+}_{\nu_1}a^{2+}_{\nu_2}\cdots a^{2+}_{\nu_{n_2}}
|0\rangle\nonumber
\ee
and the components
$\Phi_{\mu_1\mu_2\cdots\mu_{n_1}, \nu_1\nu_2\cdots\nu_{n_2}}(x)$
are automatically symmetrical under the permutations of indices of
the same type.

The conditions  \p{sym} -- \p{trans2} can be easily expressed
in this auxiliary Fock space as
\begin{eqnarray}\label{sym1}
&&T|\Phi\rangle = 0,\\
&&L^{ij} |\Phi\rangle = 0,\\
&&L^0 |\Phi\rangle = L^i |\Phi\rangle = 0,
\end{eqnarray}
where the operators
\begin{equation}
T= a^{+1}_{\mu}a^2_{\mu},\quad L^{11} = \frac{1}{2}a^1_{\mu}a^1_{\mu},\quad
L^{12} = a^1_{\mu}a^2_{\mu},\quad
L^{22} = \frac{1}{2}a^2_{\mu}a^2_{\mu},\quad
L^i = p_\mu a_{\mu}^i,
\end{equation}
along with their conjugates
\begin{equation}
T^+\!=\! a^{+2}_{\mu}a^1_{\mu},\;
L^{11 +}\!=\!\frac{1}{2}a^{1+}_{\mu}a^{1+}_{\mu},\;
L^{12 +}\!=\!a^{1+}_{\mu}a^{2+}_{\mu},\;
L^{22 +}\!=\!\frac{1}{2}a^{2+}_{\mu}a^{2+}_{\mu},\;
L^{i+}\!=\!p_\mu a_{\mu}^{i+}
\end{equation}
 and
operators
\begin{equation}
L^0 = - p^2_{\mu},\quad H^i = -a^{i+}_{\mu}a^i_{\mu} + \frac{D}{2}
\end{equation}
satisfy the following nonzero commutation relations
\begin{eqnarray}  \label{ALGEBRA}
&&[L^{11},L^{11 +}]=H^1, \quad  [L^{22},L^{22+}]=H^2,
\quad [L^{12},L^{12+}]=H^1+H^2, \nonumber \\
&&[T^+,L^{11 +}]=-L^{12 +}, \quad  [T^+,L^{12 +}]=-2L^{22+},  \nonumber \\
&&[T,L^{11}]=L^{12}, \quad  [T,L^{12}]=2L^{22},
\quad  [T,T^+]=H^1-H^2, \nonumber \\
&&[L^{22},L^{12 +}]=-T, \quad  [L^{22},T^+]=-L^{12},
\quad  [T,L^{22 +}]=-L^{12 +}, \\
&&[L^{11},L^{12 +}]=-T^+, \quad  [L^{12},L^{11+}]=-T,
\quad  [L^{12},L^{22 +}]=-T^+, \nonumber \\
&&[T,L^{12 +}]=-2L^{11 +}, \quad  [T^+,L^{12}]=2L^{11} \nonumber \
\end{eqnarray}
and
\bea
&&[L^{i}\;,\;T^+]=- \delta^{i2}L^{1},\;\;
[L^{i}\;,\;T]= - \delta^{i1} L^{2}, \nonumber \\
&&[L^{i+}\;,\;T^+]= \delta^{i1}L^{2+},\;\;
[L^{i+}\;,\;T]=  \delta^{i2} L^{1 +}, \nonumber \\
&&[L^{i}\;,\;L^{11 +}]= - \delta^{i1}L^{1+},\;\;
[L^{i+}\;,\;L^{11}]=  \delta^{i1} L^{1},  \\
&&[L^{i}\;,\;L^{22 +}]= - \delta^{i2}L^{2+},\;\;
[L^{i+}\;,\;L^{22}]=  \delta^{i2} L^{2}, \nonumber \\
&&[L^{i}\;,\;L^{12 +}]= - \delta^{i1}L^{2+} - \delta^{i2}L^{1+} ,\;\;
[L^{i+}\;,\;L^{12}]= \delta^{i1}L^{2} + \delta^{i2}L^{1},  \nonumber \\
&&[L^{i}\;,\;L^{j +}]= \delta^{ij}L^0.\nonumber
\eea

Note the asymmetry between the
operators $a^{1+}_{\mu}$ and $a^{2+}_{\mu}$ in
\p{sym1}. The later equation means
in turn that just \p{sym} is fulfilled where one index of
the second raw is symmetrized with indices of the first one.
In order
to restore the mentioned symmetry, the following equation
\begin{equation}\label{asym}
  T^+|\Phi\rangle = 0,
\end{equation}
 should be added to the system of constraints. The consistency
of the equations \p{sym1} and \p{asym}
\begin{equation}\label{conf}
[T, T^+]|\Phi\rangle =(H^1-H^2)|\Phi\rangle =(n_1-n_2)|\Phi\rangle = 0
\end{equation}
implies the equality of numbers $n_1=n_2=n$
of the operators $a^{1+}_{\mu}$ and $a^{2+}_{\mu}$
in the Fock space vector \p{vector}. However, in this case
one can show
using only  equation \p{sym1}  the following property
of the basic field
\begin{equation}\label{confsym}
  \Phi_{\mu_1\mu_2\cdots\mu_{n}, \nu_1\nu_2\cdots\nu_{n}}(x)=
(-1)^n \Phi_{\nu_1\nu_2\cdots\nu_{n}, \mu_1\mu_2\cdots\mu_{n}}(x)
\end{equation}
and, as a consequence, the equation \p{asym} is also
fulfilled. Moreover,
there is no nonzero solution
 of the equation \p{sym1}
for the case of $n_1<n_2$. All this  means that the
Fock space vector  under the constraint \p{sym1}
contains  all possible Young tableaux
of the type \p{young}
and only once each of them.

The set  of operators $L^{ij}, T, L^{ij +}, T^+, H^i$
form the algebra \p{ALGEBRA} of the group $SO(3,2)$,
$H^i$  being the Cartan generators.
Since the conditions $H^i|\Phi\rangle  = 0$ can not be satisfied
for any nonzero vector $|\Phi\rangle$, the generators $H^i$
must be excluded from the total set of constraints, and therefore
 one obtains the system of the first and the second class constraints.
The corresponding nilpotent BRST charge can be constructed as follows
\cite{BPT}.

The first step is to construct the auxiliary representations
of the algebra  $SO(3,2)$ using the Verma module
after introduction of the additional creation and annihilation
operators $\left[ b_I,b^{+}_J \right] = \delta_{IJ}$,
$I,J=1,...,4$
(The general method for such constructions
was given in \cite{B}). Note that the number of
the oscillators is equal to the number of positive roots of
algebra $SO(3,2)$ and the vector $|\Phi\rangle$
depends also on the creation operators $b^+_I$.  The auxiliary
representation  has the form:
\begin{eqnarray} \label{aux}
&&L^{11+}_{aux.}=b_1^+, \quad L^{12+}_{aux.}=b_2^+,
\quad L^{22+}_{aux.}= b_3^+, \nonumber \\
&&T^+_{aux.}=b_4^+-b_2^+b_1-2b_3^+b_2, \nonumber \\
&&T_{aux.}=(h_1-h_2-b_4^+b_4)b_4-2b_1^+b_2-b_2^+b_3, \nonumber \\
&&H^1_{aux.}=2b_1^+b_1+b_2^+b_2-b_4^+b_4+h_1 \nonumber, \\
&&H^2_{aux.}=b_2^+b_2+2b_3^+b_3+b^+_4b_4+h_2, \\
&&L^{11}_{aux.}=(b_1^+b_1+b_2^+b_2-b_4^+b_4+h_1)b_1-b_4^+b_2+b_3^+b_2b_2, \nonumber \\
&&L^{12}_{aux.}=(2b_1^+b_1+b_2^+b_2+2b_3^+b_3+h_1+h_2)b_2 + b^+_4b_4b_4b_1 \nonumber \\
&&\hskip1cm +b_2^+b_3b_1-b_4^+b_3 + (h_2-h_1)b_4b_1, \nonumber \\
&&L^{22}_{aux.}=(b_2^+b_2+b_3^+b_3+b_4^+b_4+h_2)b_3 \nonumber \\
&&\hskip1cm +(h_2-h_1)b_4b_2
+b_1^+b_2b_2+b_4^+b_4b_4b_2, \nonumber \
\end{eqnarray}
where $h_1$ and $h_2$ are parameters and the auxiliary
representations of operators $H^i$ depend on them linearly.

Further, let us denote ${E}^{\alpha} \equiv (L^{ij}, T)\quad (\alpha>0)$,
and define
\begin{equation} \label{sum}
{\cal H}^i={H}^i + \tilde {H^i}_{aux.} +  h^i ,\quad
{\cal E}^{\alpha}= {E}^{ \alpha} + E_{aux.}^{ \alpha}(h),
\end{equation}
where we have explicitly extracted the dependence on parameters $h_i$
in the auxiliary representations of Cartan generators.
 After writing the $SO(3,2)$ algebra in the compact form
\begin{eqnarray}
\label{commutator}
&&\left[{\cal H}^i,{\cal E}^\alpha\right]=\alpha(i) {\cal E}^\alpha, \nonumber\\
&&\left[{\cal E}^\alpha,{\cal E}^{-\alpha}\right]=\alpha^i {\cal H}^i,\\
&&\left[{\cal E}^\alpha,{\cal E}^{\beta}\right]=N^{\alpha\beta}
{\cal E}^{\alpha+\beta}, \nonumber
\end{eqnarray}
 we introduce the anticommuting variables
$\e_\alpha \equiv
(\e_{ij},\e_{T})$ $\e_{-\alpha}=\e_\alpha^+$, having ghost number one
and corresponding momenta
$\cp_{-\alpha}=\cp_\alpha^+$ and $\cp_\alpha$,  having ghost number
minus one with the
commutation relations:
\begin{equation}
\{\e_\alpha,\cp_{-\beta}\}=
\{\e_{-\alpha},\cp_\beta\}=\delta_{\alpha\beta}.
\end{equation}
The ``ghost vacuum" is defined  as
\begin{equation}
\e_\alpha|0\rangle=\cp_\alpha|0\rangle =0
\end{equation}
for positive roots $\alpha$. The nilpotent BRST charge
for the above subsystem
of constraints with no $H^i$ dependence has the form

\begin{equation}
{Q}_1=\sum_{\alpha>0}\left(\e_\alpha {\cal E}^{-\alpha}+
\e_{-\alpha}{\cal E}^{\alpha}\right)-
\frac{1}{2}\sum_{\alpha\beta}N^{\alpha\beta}
\e_{-\alpha}\e_{-\beta}\cp_{\alpha+\beta}
\end{equation}
where the parameters $h^i$ have to be substituted by the
expressions \cite{BPT}
\begin{equation}
-\pi^{i}=
-{ H}^{i} -\tilde{H}^{i}_{aux.}-
\sum_{\beta>0}\beta(i)\left(\e_\beta\cp_{-\beta}-
\e_{-\beta}\cp_\beta\right)
\end{equation}

The inclusion of the constraints $\cal{L}^A$ $\equiv (L^0, L^i, L^{i+})$
into the total BRST charge $Q$ is trivial, namely
\begin{equation} \label{BRST}
Q=Q_1 + Q_2
\end{equation}
where
\begin{equation}
Q_2 = \e_0 L^0 + \e_i L^{i +} + \e^{+}_iL^i -
\e^{ +}_i \e_i \cp_0
+ \sum_{\alpha > 0, A,B}(\e_A \e^{+}_\alpha \cp_B C_{- \alpha, A}^B
+  \e_A \e_{\alpha} \cp_B C_{ \alpha, A}^B)
\end{equation}
in self explanatory notations. This completes the procedure
of constructing nilpotent BRST charge for our system.

\setcounter{equation}0\section{The lagrangian and
the partial gauge fixing}

The BRST invariant lagrangian which describes the
massless irreducible  representations of the Poincare group of the form
\p{young} can be written as
\begin{equation} \label{L}
- L = \int d \e_0 \langle \chi| K Q | \chi \rangle,
\end{equation}
being invariant under the gauge transformations
\begin{equation} \label{G}
\delta | \chi \rangle = Q | \Lambda \rangle
\end{equation}
with a parameter of gauge transformations $| \Lambda \rangle$.
The Kernel operator $K$  in the scalar product \p{L}
is necessary to make the lagrangian hermitian \cite{BPT},
since the BRST charge Q is not hermitian as one can conclude from
the explicit form of the auxiliary representations \p{aux}.
The operator $K$ is constructed as follows.
Let us introduce the vector in the  space of Verma module
\begin{equation}\label{b1}
{\left|n_1,n_2,n_3,n_4 \right \rangle}_V =
(L^{11 +})^{n_1}(L^{12+})^{n_2}(L^{22+})^{n_3}
 (T^+)^{n_4}{|0 \rangle}_V
\end{equation}
where ${\alpha_1}, {\alpha_2}, \dots {\alpha_r}$
is some ordering of positive roots,
 $n_i \in N$ and $E^{\alpha}{|0 \rangle}_V = 0$
The corresponding vector in the Fock space generated by the creation and
annihilation operators $b_I, b_I^+$ is
  \begin{equation}\label{b2}
\left|n_1,n_2,n_3,n_4 \right\rangle =
(b^+_1)^{n_1}(b^+_2)^{n_2}(b^+_3)^{n_3} (b^+_4)^{n_4}|0 \rangle.
\end{equation}
The Kernel operator $K$   defining
the scalar product of two vectors $|\Phi_1\rangle$
and $|\Phi_2\rangle$ as $\langle\Phi_2|K|\Phi_1\rangle$
has the form
\begin{equation}
K = Z^+ Z,
\end{equation}
where the operator $Z$ transforms the given state from the Fock space
to the corresponding state in the Verma module and have the following form
\begin{equation}
Z=
\sum_{n_i} \frac{1}{\prod n_i!}
(L^{11 +})^{n_1}(L^{12+})^{n_2}(L^{22+})^{n_3}
 (T^+)^{n_4}{|0 \rangle}_V
 \langle 0|(b_1)^{n_1}(b_2)^{n_2}(b_3)^{n_3} (b_4)^{n_4}.
\end{equation}
Such form of the operators $Z$ and $K$ guarantees the identity of
the scalar products of the corresponding vectors in the Fock space and in
the Verma module.
Then the following modified hermiticity relation is satisfied
\cite{BPT}
\begin{equation}
Q^+ K = K Q
\end{equation}

In order to be physical the lagrangian \p{L} must have ghost number
equal to zero and therefore the ghost number of the vector
$| \chi \rangle$ which in turn is the series
expansion with respect to creation operators
$\e_0,\e^{+}_i, \cp^{+}_i, \e^+_{\alpha }, \cp^+_{\alpha}$, along with
operators $a_\mu^{i +}$ and $b_I^+$
  must be zero as well.
The same applies to the parameter of the gauge transformations
$| \Lambda \rangle$ which has the ghost number equal to $-1$.

Below we prove, following
a slightly different way than in \cite{PT2},
that the vector $| \chi \rangle$ contains the only physical
field $| S_0 \rangle$ with no $b^+_I$ dependence. The other fields
can be either excluded using the equations of motion or gauged away.

First let us write explicitly the $\cp_\alpha^+$ dependence of the
vector  $| \chi \rangle$    and the parameter $| \Lambda \rangle$
\begin{equation} \label{expans}
| \chi \rangle = | \chi_0 \rangle + \cp_\alpha^+| \chi_\alpha \rangle +
\cp_\alpha^+ \cp_\beta^+| \chi_{\alpha \beta} \rangle + ...
\end{equation}
\begin{equation}
| \Lambda \rangle = | \Lambda_0 \rangle + \cp_\alpha^+|
\Lambda_\alpha \rangle +
\cp_\alpha^+ \cp_\beta^+| \Lambda_{\alpha \beta} \rangle + ...
\end{equation}
From the gauge transformation law \p{G} and the explicit form of the
auxiliary representations of $SO(3,2)$ algebra \p{aux} one can see
that the field  $| \chi_0 \rangle$    transforms through parameters
$|\Lambda_\alpha \rangle$
as
\begin{eqnarray}
\delta| \chi_0 \rangle &=&  (L^{11 +} + b_1^+)| \Lambda_{11} \rangle
+ (L^{12 +} + b_2^+)| \Lambda_{12} \rangle \nonumber \\
&&+
(L^{22 +} + b_3^+)| \Lambda_{22} \rangle   +
(T^{ +} + b_4^+ - b_2^+b_1 - 2b_3^+ b_2)| \Lambda_{T} \rangle.
\end{eqnarray}
As it was shown in \cite{PT2} one can use this gauge freedom to gauge away
all $b_I^+$ dependence in $| \chi_0 \rangle$ and therefore
\begin{equation} \label{NB}
b_I| \chi_0 \rangle = 0
\end{equation}
Let us note, that one is left with the residual gauge freedom
in transformations of $| \chi_0 \rangle$ with the parameter
$\Lambda_{0} \rangle$. However this parameter is irrelevant
to the proof of $b_I^+$ independence of the vector
$| \chi_0 \rangle$.

As the second step we use the ideology of auxiliary BRST conditions
\cite{M}. Namely since the equations of motion resulting from the
lagrangian \p{L} have the form\footnote{Since the Kernel operator
$K$ is nondegenerate one can multiply the equation
$KQ| \chi \rangle = 0$ on $K^{-1}$ to obtain \p{EOM}}
\begin{equation} \label{EOM}
Q| \chi \rangle = 0
\end{equation}
one can also impose on $| \chi \rangle$
the auxiliary conditions
\begin{equation} \label{CA}
R_I| \chi \rangle = \left[M_I, Q \right]_{\pm}| \chi \rangle
= Q M_I  | \chi \rangle =0
\end{equation}
for some operators $M_I$. Such conditions do not affect the physical content
of the theory since they remove the states with the zero norm.
The choice of the operators $M_I$ is actually the choice of the BRST gauge.

Taking $M_I = b_I$, one obtains
\begin{equation} \label{UR}
(\e_\alpha + A_\alpha(\e^+_{\alpha}, b))| \chi \rangle =0
\end{equation}
where the  operators $A_{\alpha}$ are written in the Appendix. Their
explicit form leads to the following solution of equations \p{UR}
\begin{equation}
| \chi \rangle = (1 + \cp_{11}^+ A_{11})
(1 + \cp_{12}^+ A_{12})
(1 + \cp_{22}^+ A_{22})(1 + \cp_{T}^+ A_{T})| \chi_0 \rangle,
\label{SOL}
\end{equation}
and from  \p{NB} and \p{A1}--\p{A4} one obtains
\begin{equation} \label{CHI}
| \chi \rangle = | \chi_0 \rangle.
\end{equation}
Until now we have proven that the all $b_I^+$ and $\cp_\alpha^+$
dependence in the vector $| \chi \rangle$   is the BRST gauge artifact.
However one has to keep in mind the equations of motion
resulting from the lagrangian \p{L} after the variation
with respect to this ``pure gauge" fields (see the Appendix).

Inserting the vector $| \chi \rangle$ of the form \p{CHI}
into  \p{L} and performing explicit calculation one
obtains  that the part $| \chi \rangle$ which contains $\e^+_{\alpha}$
dependence completely decouples from the lagrangian,
therefore the state vector has effectively the form
\begin{eqnarray} \label{vector}
|\chi\rangle&=&|S_0\rangle +
{\eta}_1^+ {\cal P}_1^+ |S_1\rangle +
{\eta}_2^+ {\cal P}_2^+ |S_2\rangle +
{\eta}_1^+ {\cal P}_2^+ |S_3\rangle
\nonumber \\
&&+{\eta}_2^+ {\cal P}_1^+ |S_4\rangle +
{\eta}_1^+ {\eta}_2^+ {\cal P}_1^+ {\cal P}_2^+ |S_5\rangle +
\eta_0 {\cal P}_1^+ |R_1\rangle  \nonumber\\
&&+\eta_0 {\cal P}_2^+ |R_2\rangle +
\eta_0 {\eta}_1^+ {\cal P}_1^+ {\cal P}_2^+ |R_3\rangle +
\eta_0 {\eta}_2^+ {\cal P}_1^+ {\cal P}_2^+ |R_4\rangle
\end{eqnarray}
with vectors $\ra{i}$ and $|R_i\rangle$
having ghost number zero and depending only
on bosonic creation operators $a_\mu^{i+}$.

Finally using the residual gauge freedom
with the parameter $|\Lambda_0 \rangle$
and equations of motion
one can gauge away the field $|S_3\rangle-|S_4\rangle$,
 express the other fields in terms of the single field
$|S_0\rangle$  and put them into the lagrangian.
The final expression for the lagrangian describing all massless
irreducible representations of the Poincare group
with the corresponding Young tableaux having two rows has the form
\begin{eqnarray} \label{F}    \nonumber
-L &=& \la{0} L^{0}  -  L^{+1}  L^{1}  -
       L^{+2} L^{2}  -
       L^{+1}  L^{+1}  L^{11} \nonumber \\
&&- L^{+1}  L^{+2}  L^{12}   -
        L^{+2}  L^{+2}  L^{22}   -
      2   L^{+11}  L^{0}  L^{11}
- L^{+11}  L^{1}  L^{1}  \nonumber \\
&& - L^{+12}  L^{0}  L^{12}   -
        L^{+12}  L^{1}  L^{2}   -
      2   L^{+22}  L^{0}  L^{22}   -
        L^{+22}  L^{2}  L^{2}  \nonumber \\
&& - L^{+1}  L^{+11}  L^{1}  L^{11}
- L^{+1}  L^{+12}  L^{2}  L^{11}   +
        L^{+1}  L^{+22}  L^{1}  L^{22} \nonumber \\
&& - L^{+1}  L^{+22}  L^{2}  L^{12}   -
        L^{+2}  L^{+11}  L^{1}  L^{12}
+ L^{+2}  L^{+11}  L^{2}  L^{11}  \nonumber \\
&&-L^{+2}  L^{+12}  L^{1}  L^{22}   -
        L^{+2}  L^{+22}  L^{2}  L^{22}   +
        L^{+1}  L^{+1}  L^{+22}  L^{11}  L^{22}  \nonumber \\
&&-  L^{+1}  L^{+2}  L^{+12}  L^{11}  L^{22}   +
        L^{+2}  L^{+2}  L^{+11}  L^{11}  L^{22}   +
      3   L^{+11}  L^{+22}  L^{0}  L^{11}  L^{22} \nonumber \\
&&+ L^{+11}  L^{+22}  L^{1}  L^{1}  L^{22}
- L^{+11}  L^{+22}  L^{1}  L^{2}  L^{12}   +
        L^{+11}  L^{+22}  L^{2}  L^{2}  L^{11}  \nonumber \\
&&+ L^{+1}  L^{+11}  L^{+22}  L^{1}  L^{11}  L^{22}   +
        L^{+2}  L^{+11}  L^{+22}  L^{2}  L^{11}  L^{22} \ra{0}
\end{eqnarray}
where  the field $\ra{0}$ is constrained as
\begin{equation} \label{USL1}
T\ra{0}=0
\end{equation}
\begin{eqnarray}
L^{ 11}L^{11}\ra{0}&=&L^{ 11}L^{12}\ra{0}=L^{ 22}L^{22}\ra{0} =
L^{ 12}L^{22}\ra{0} \nonumber\\
=(L^{ 12}L^{12} +2L^{ 11}L^{22})\ra{0} &=&0.\label{USL2}
\end{eqnarray}

From the algebra \p{ALGEBRA} one can conclude that the constraints
given above
 are consistent with each other. Actually there are two independent
constraints  on the basic field, namely the constraint
  \p{USL1} and first of the constraints \p{USL2},
 the other ones can be considered as the consistency conditions
for this system.

The lagrangian \p{F} is invariant under the transformations
\begin{equation}
\delta \ra{0} = L^{i+} |\lambda_i\rangle \quad i =1,2.
\end{equation}
parameters $|\lambda_i\rangle$
coming from $\cp^+_1|\lambda_1\rangle +\cp^+_2|\lambda_2\rangle$
term in $|\Lambda_0\rangle$
are constrained as
\begin{equation}\label{gfconserv}
L^{kl}|\lambda_i\rangle= T|\lambda_2\rangle=0 \quad
 T|\lambda_1\rangle =|\lambda_2\rangle
\end{equation}
The conditions \p{gfconserv} are necessary to maintain the
gauge fixed form \p{vector}
(or equivalently the gauge invariance of  constraints
\p{USL1} --\p{USL2})
 of the wavefunction $|\chi\rangle$
with respect the residual gauge transformations \p{G}.
Finally one arrives at the transformations with the single gauge
parameter  $|\lambda\rangle=|\lambda_1\rangle$
\begin{equation}
\delta \ra{0} = (L^{1+} +L^{2+}T)|\lambda\rangle
\end{equation}
constrained as follows
\begin{equation}\label{constraintsforlambda}
L^{ij}|\lambda\rangle = T^2 |\lambda\rangle =0
\end{equation}

\setcounter{equation}0\section{Examples}
In this section we construct
the explicit form of  the lagrangians for some
simple Young tableaux which correspond to lower orders in the expansion
of the field $\ra{0}$.
\begin{itemize}
\item
\begin{picture}(20,20)
\unitlength=0.5mm
\put(0,7.5){\line(1,0){5}}
\put(0,2.5){\line(1,0){5}}
\put(0,-2.5){\line(1,0){5}}
\put(0,-2.5){\line(0,1){10}}
\put(5,-2.5){\line(0,1){10}}
\end{picture}:
 $\ra{0}=\Phi_{\mu,\nu}(x)a^{1+}_{\mu}a^{2+}_{\nu}|0\rangle$\vspace{5mm}\\
The  equation \p{USL1} means antisymmetry of the
 field: $\Phi_{\mu,\nu}(x)=-\Phi_{\nu,\mu}(x)$.
The lagrangian  \p{F} for this antisymmetric field
 (the notoph in four dimensions \cite{OP})
\begin{equation}\label{L11}
  L=-\Phi_{\mu,\nu}\partial^2_\rho\Phi_{\mu,\nu}+
2\Phi_{\mu,\nu}\partial_{\mu}\partial_{\rho}
\Phi_{\rho,\nu}
\end{equation}
in terms of the field strength
$F_{\mu\nu\rho}=\partial_\mu\Phi_{\nu,\rho}+
\partial_\nu\Phi_{\rho,\mu}+\partial_\rho\Phi_{\mu,\nu}$
has the standard form
\begin{equation}\label{LS11}
  L=\frac{1}{3}F_{\mu\nu\rho}^2
\end{equation}
and is invariant under the well known gauge transformations
\begin{equation}\label{G11}
  \delta\Phi_{\mu,\nu}(x)=\partial_\mu\lambda_\nu(x)-
\partial_\nu\lambda_\mu(x).
\end{equation}

\item
\begin{picture}(20,20)
\unitlength=0.5mm
\put(0,7.5){\line(1,0){10}}
\put(0,2.5){\line(1,0){10}}
\put(0,-2.5){\line(1,0){5}}
\put(0,-2.5){\line(0,1){10}}
\put(5,-2.5){\line(0,1){10}}
\put(10,2.5){\line(0,1){5}}
\end{picture}:
 $\ra{0}=\Phi_{\mu\nu,\rho}(x)a^{1+}_{\mu}a^{1+}_{\nu}
a^{2+}_{\rho}|0\rangle$\vspace{5mm}\\
The symmetry   with respect to the first two indices
$\Phi_{\mu\nu,\rho}=\Phi_{\nu\mu,\rho}$
which is guaranteed by the construction and the condition
\p{USL1} lead to the following property of
the field  $\Phi_{\mu\nu,\rho}$:
\begin{equation}\label{S21}
 \Phi_{\mu\nu,\rho}+\Phi_{\rho\nu,\mu}+\Phi_{\mu\rho,\nu}=0.
\end{equation}
Taking all this into account the
lagrangian \p{F} for the third rank tensor
field $\Phi_{\mu\nu,\rho}$ corresponding to the
considered Young tableaux can
be written in the form
\begin{eqnarray}\label{L21}
  L&=&2 \Phi_{\mu\nu,\rho}\partial^2_\sigma \Phi_{\mu\nu,\rho}-
3\Phi_{\mu\mu,\rho}\partial^2_\sigma \Phi_{\nu\nu,\rho}-
4\Phi_{\mu\nu,\rho}\partial_\mu\partial_\sigma \Phi_{\sigma\nu,\rho}-
2 \Phi_{\mu\nu,\rho}\partial_\rho\partial_\sigma
\Phi_{\mu\nu,\sigma} \nonumber\\
&&+6\Phi_{\mu\nu,\rho}\partial_\mu\partial_\nu \Phi_{\sigma\sigma,\rho}+
3\Phi_{\mu\mu,\rho}\partial_\rho\partial_\sigma \Phi_{\nu\nu,\sigma},
\end{eqnarray}
and is invariant under the following gauge transformations
\begin{equation}\label{G21}
  \delta\Phi_{\mu\nu,\rho}(x)=\partial_\mu\lambda_{\nu,\rho}(x)+
\partial_\nu\lambda_{\mu,\rho}(x)-
\partial_\rho\lambda_{\mu,\nu}(x)-
\partial_\rho\lambda_{\nu,\mu}(x),
\end{equation}
the gauge transformation parameter being traceless
$\lambda_{\mu,\mu}=0$.

\item
\begin{picture}(20,20)
\unitlength=0.5mm
\put(0,7.5){\line(1,0){10}}
\put(0,2.5){\line(1,0){10}}
\put(0,-2.5){\line(1,0){10}}
\put(0,-2.5){\line(0,1){10}}
\put(5,-2.5){\line(0,1){10}}
\put(10,-2.5){\line(0,1){10}}
\end{picture}:
 $\ra{0}=\Phi_{\mu\nu,\rho\sigma}(x)a^{1+}_{\mu}a^{1+}_{\nu}
a^{2+}_\rho a^{2+}_\sigma |0\rangle$\vspace{5mm}\\
The field $\Phi_{\mu \nu, \rho \sigma}$ is symmetrical with respect
to the permutations  of the indices $(\mu , \nu)$ and $(\rho, \sigma)$
by the construction and
obeys the   relations
\begin{equation}\label{S22}
 \Phi_{\mu\nu,\rho \sigma}+\Phi_{\rho\nu,\mu \sigma}+
\Phi_{\mu\rho,\nu \sigma}=0
\end{equation}
and
\begin{equation}
 \Phi_{\mu\mu,\rho\rho}=-2\Phi_{\mu \rho,\mu\rho}
\end{equation}
which is the only nontrivial among the constraints \p{USL1}
as a consequence of \p{S22}. Moreover,  according  to   \p{confsym}
it is also symmetrical under the
permutation of pairs of indices
\begin{equation}\label{confsym22}
  \Phi_{\mu\nu,\rho\sigma}=\Phi_{\rho\sigma,\mu\nu}.
\end{equation}

The lagrangian for this field has the form
\begin{eqnarray}\label{L22}
  L&=&-4 \Phi_{\mu\nu,\rho \tau}\partial^2_\sigma \Phi_{\mu\nu,\rho\tau}
+12\Phi_{\mu\nu,\rho \rho }\partial^2_\sigma\Phi_{\mu\nu,\tau \tau }
-3\Phi_{\mu\mu,\nu \nu}\partial^2_\rho \Phi_{\sigma \sigma,\tau \tau}
\nonumber \\
&&+16\Phi_{\mu\nu,\rho \sigma}\partial_\mu\partial_\tau
 \Phi_{\tau \nu,\rho \sigma}
-24\Phi_{\mu\nu,\rho \sigma}\partial_\mu \partial_\nu
 \Phi_{\tau\tau,\rho \sigma}
  \nonumber \\
&& -24\Phi_{\mu\nu,\rho \rho}\partial_\mu \partial_\sigma
 \Phi_{\sigma\nu,\tau \tau}
+12\Phi_{\mu\nu,\rho \rho}\partial_\mu \partial_\nu
 \Phi_{\sigma\sigma,\tau \tau}
\end{eqnarray}
and is invariant under the transformations
\begin{equation}
\delta\Phi_{\mu\nu,a b}(x)=\partial_\mu \lambda_{\nu,ab}(x)
- 2 \partial_a \lambda_{\mu,\nu b}(x)
\end{equation}
where the symmetrization over couples of Greek and Latin
indices is assumed. The parameter of the gauge transformations
$\lambda_{\mu, \nu \rho}$ is symmetrical with respect to the indices
$\nu$ and $\rho$ and obeys the constraints \p{constraintsforlambda}
\begin{equation}
\lambda_{\mu,\mu \nu}= \lambda_{\mu,\nu \nu}=0,
\end{equation}
\begin{equation}
\lambda_{\mu,\nu\rho}+ \lambda_{\nu,\rho\mu}+\lambda_{\rho,\mu\nu}=0
\end{equation}and, therefore it is described by the Young tableaux of type
\begin{picture}(20,20)
\unitlength=0.5mm
\put(0,7.5){\line(1,0){10}}
\put(0,2.5){\line(1,0){10}}
\put(0,-2.5){\line(1,0){5}}
\put(0,-2.5){\line(0,1){10}}
\put(5,-2.5){\line(0,1){10}}
\put(10,2.5){\line(0,1){5}}
\end{picture}.
\end{itemize}

Although the symmetry properties of the tensor $\Phi_{\mu \nu, \rho \sigma}$
do  not coincide with those for Weyl tensor $C_{\mu \nu, \rho \sigma}$
\begin{equation}
C_{\mu \nu, \rho \sigma} = - C_{\nu \mu, \rho \sigma}=-C_{\mu \nu, \sigma
\rho},
\quad C_{\mu \nu, \rho \sigma} = C_{\rho \sigma, \mu \nu},
\end{equation}
they can be
related to each other with the help of the following transformations
\begin{eqnarray}
\Phi_{\mu \nu, \rho \sigma} &=& \frac{1}{4}
(C_{\mu \rho, \nu \sigma} + C_{\mu \sigma, \nu \rho})\\
C_{\mu \rho, \nu \sigma} &=& \frac{4}{3}(\Phi_{\mu \nu, \rho \sigma}-
\Phi_{\rho \nu, \mu \sigma}).
\end{eqnarray}
Strictly speaking $C_{\mu \nu, \rho \sigma}$  is not a Weyl tensor because its traces
do not vanish. However,
one can obtain  from \p{L22}  the following lagrangian
\begin{eqnarray}\label{W}
  L&=&-\frac{1}{2}
C_{\mu\rho,\nu \tau}\partial^2_\sigma C_{\mu\rho,\nu\tau}
-\frac{1}{2}
C_{\mu\tau,\nu \rho}\partial^2_\sigma C_{\mu\rho,\nu\tau}
+ 3
C_{\mu\rho,\nu \rho}\partial^2_\sigma C_{\mu\tau,\nu\tau} \nonumber \\
&& - \frac{3}{4}
C_{\mu\nu,\mu \nu}\partial^2_\sigma C_{\rho\tau,\rho\tau}
+2
C_{\mu\rho,\nu \sigma}\partial_\mu \partial_\tau
 C_{\tau\rho,\nu\sigma}
+2
C_{\mu\rho,\nu \sigma}\partial_\mu \partial_\tau
 C_{\tau\sigma,\nu\rho} \nonumber \\
&&
-6
C_{\mu\rho,\nu \sigma}\partial_\mu \partial_\nu
 C_{\tau\rho,\tau\sigma}
-6
C_{\mu\rho,\nu \rho}\partial_\mu \partial_\sigma
 C_{\sigma\tau,\nu\tau}
+3
C_{\mu\rho,\nu \rho}\partial_\mu \partial_\nu
 C_{\sigma\tau,\sigma\tau}
\end{eqnarray}
from which  the vanishing of all traces
of this tensor on mass shell follows. Therefore one can conclude
 that the lagrangian \p{W} consistently
describes the free field theory of the Weyl tensor.

\setcounter{equation}0\section{Conclusions}
In the present paper we have explicitly constructed
the lagrangians  describing massless
irreducible representations of
the Poincare group of the form \p{young}.
 As it was mentioned in the
introduction and is clear from the
calculations above the approach
used for this construction can be directly applied for
the lagrangian description of an arbitrary
representation of the Poincare
group as well.

It seems to be interesting to generalize this technique also
for the description of the interaction of
higher spin fields
with some gravitational background. This will obviously lead
to the modification of the system of
constraints present in the BRST
charge. The problem of constructing of the nilpotent BRST charge for
this kinds of physical systems can in
turn reveal an allowed
types of gravitational backgrounds where  the higher spin
fields can propagate consistently.

\vspace{0.3cm}

\noindent {\bf Acknowledgments}
~This ~investigation ~has ~been ~supported
~by ~the
~grant ~of ~the ~Committee ~for collaboration between Czech Republic and JINR.
The work of \v{C}.B. was supported by the grant of the Ministry
of Education of Czech Republic VZ/400/00/18. The work of A.P. and M.T.
was supported
 in part by the
Russian Foundation of Basic Research,
grant 99-02-18417 and the joint grant RFBR-DFG
99-02-04022.

\vspace{1cm}

\noindent {\Large\bf Appendix}
\setcounter{equation}0
\def\theequation{A\arabic{equation}}

\vspace{.5cm}

In this Appendix we present some formulae
implicitly used in the main body
of the paper.

The explicit form of the
operators $A_\alpha(\e^+_{\alpha}, b)$ in \p{UR} shows that they indeed
depend only on the operators $\e^+_{\alpha}$ and $b_I$ --
 the property crucial
for the establishing of the relation \p{CHI} :

\begin{eqnarray} \label{A1}
A_{11}&=& -2 \e^+_T  b_2 - 2 \e^+_T  b_1  b_4 - \e^+_{11}  b_1  b_1 +
  \e^+_{22}  b_2  b_2
+ 2 \e^+_{12}  b_1  b_1  b_4 \nonumber \\
&&+
  2 \e^+_{22}  b_1  b_2  b_4,\\
A_{12}&=&-\e^+_T  b_3 - \e^+_{11}  b_1  b_2 - \e^+_{12}  b_2  b_2 +
  \e^+_T  b_1  b_4  b_4
- \e^+_{12}  b_1  b_1  b_4  b_4 \nonumber \\
&&-
  \e^+_{22}  b_1  b_2  b_4  b_4,\\
A_{22}&=& 2 \e^+_T  b_3  b_4 - \e^+_{11}  b_2  b_2 - 2 \e^+_{12}  b_2  b_3 -
   \e^+_{22}  b_3  b_3
+ 2 \e^+_T  b_2  b_4  b_4 \nonumber \\
&& -2 \e^+_{12}  b_1  b_3  b_4 - 2 \e^+_{22}  b_2  b_3  b_4 -
  2 \e^+_{12}  b_1  b_2  b_4  b_4 -
  2 \e^+_{22}  b_2  b_2  b_4  b_4,\\
\label{A4}
A_T&=& - \e^+_{11}  b_2 - \e^+_{12}  b_3 + \e^+_T  b_4  b_4
-\e^+_{12}  b_1  b_4  b_4 - \e^+_{22}  b_2  b_4  b_4.
\end{eqnarray}

The lagrangian \p{L} after using the explicit expression
of the state vector \p{vector} and  of the BRST charge \p{BRST}
takes the form
\begin{eqnarray}  \label{LLL}
-L& =&
    \langle R_1|  |R_1 \rangle + \langle R_2|  |R_2 \rangle -
     \langle R_3|  |R_3 \rangle - \langle R_4|  |R_4 \rangle
- \langle R_1|  L^{1}  \ra{0} \nonumber \\
&&+ \langle R_1|  L^{+1}  \ra{1} +
      \langle R_1|  L^{+2}  \ra{3} - \langle R_2|  L^{2}  \ra{0} +
      \langle R_2|  L^{+1}  \ra{4}  \nonumber \\
&&+ \langle R_2|  L^{+2}  \ra{2}
- \langle R_3|  L^{1}  \ra{4} + \langle R_3|  L^{2}  \ra{1} +
      \langle R_3|  L^{+2}  \ra{5}  \nonumber \\
&&-\langle R_4|  L^{1}  \ra{2}
+ \langle R_4|  L^{2}  \ra{3}
-\langle R_4|  L^{+1}  \ra{5} + \la{0}  L^{0}  \ra{0} \nonumber \\
&&- \la{0}  L^{+1}  |R_1 \rangle - \la{0}  L^{+2} |R_2 \rangle
- \la{1}  L^{0}  \ra{1}
+ \la{1}  L^{1}  |R_1 \rangle    \nonumber \\
&&+
      \la{1}  L^{+2}  |R_3 \rangle -
       \la{2}  L^{0}  \ra{2} + \la{2}  L^{2}  |R_2 \rangle
- \la{2}  L^{+1}  |R_4 \rangle  \nonumber \\
&&- \la{3}  L^{0}  \ra{4} + \la{3} L^{2}  |R_1 \rangle +
       \la{3}  L^{+2}  |R_4 \rangle - \la{4}  L^{0}  \ra{3}  \nonumber \\
&&+ \la{4}  L^{1}  |R_2 \rangle
- \la{4}  L^{+1}  |R_3 \rangle + \la{5}
      L^{0}  \ra{5}     \nonumber \\
&& -
       \la{5}  L^{1}  |R_4 \rangle + \la{5}  L^{2}  |R_3 \rangle.
 \end{eqnarray}

The equations of motion resulting from the lagrangian
\p{LLL} are
\begin{eqnarray}                     \label{equation1}
S0:&& L^{+1} | R_1 \rangle + L^{+2} | R_2 \rangle - L^{0} \ra{0}= 0,\nonumber \\
S1:&&  L^{1} | R_1 \rangle + L^{+2} | R_3 \rangle -
L^{0} \ra{1}= 0,\nonumber\\
S2:&&  L^{2} | R_2 \rangle - L^{+1} | R_4 \rangle -
L^{0} \ra{2}= 0,\nonumber\\
S3:&&  L^{2} | R_1 \rangle + L^{+2} | R_4 \rangle -
L^{0} \ra{4}= 0,\nonumber\\
S4:&&  L^{1} | R_2 \rangle - L^{+1} | R_3 \rangle -
L^{0} \ra{3}= 0,\nonumber\\
S5:&&  L^{2} | R_3 \rangle - L^{1} | R_4 \rangle +
L^{0} \ra{5}= 0,\nonumber\\
R1:&&  | R_1 \rangle - L^{1} \ra{0} + L^{+1} \ra{1} +
L^{+2} \ra{3}= 0,\nonumber\\
R2:&&  | R_2 \rangle - L^{2} \ra{0} + L^{+2} \ra{2} +
L^{+1} \ra{4}= 0,\nonumber\\
R3:&&  | R_3 \rangle - L^{2} \ra{1} + L^{1} \ra{4} -
L^{+2} \ra{5}= 0,\nonumber\\
R4:&&  | R_4 \rangle + L^{1} \ra{2} - L^{2} \ra{3} +
L^{+1} \ra{5}= 0,
\end{eqnarray}
while the  equations of motion obtained from the
variation of the initial lagrangian \p{L}
with respect to other fields
in the expansion \p{expans} which couple to the $| \chi_0 \rangle$
look as follows

\begin{equation}\label{equation2}
{\begin{tabular}{lll}
$  T \ra{0}= 0,  $ & $
  T | R_1 \rangle - | R_2 \rangle= 0,  $ & $
 T | R_2 \rangle= 0,
$\\
&&\\
$  T \ra{1} - \ra{3} - \ra{4}= 0,     $ & $
  \ra{2} - T \ra{3}= 0,    $ & $
  L^{11} \ra{1}= 0,
$\\&&\\
$
  L^{11} \ra{3}= 0,  $ & $
   L^{12} \ra{3} + \ra{5}= 0,  $&$
L^{12} \ra{1}= 0,
$\\&&\\$
  L^{22} \ra{1} - \ra{5}= 0,   $&$                                                                  \
  \ra{2} - T \ra{4}= 0, $&$
  L^{22} \ra{3}= 0,
$\\&&\\$
  T \ra{2}= 0,   $&$
  L^{11} \ra{2} - \ra{5}= 0, $&$
  L^{11} \ra{4}= 0,
$\\&&\\$
  L^{12} \ra{4} + \ra{5}= 0,  $&$
  L^{12} \ra{2}= 0,  $&$
  L^{22} \ra{4}= 0,
$\\&&\\$
  L^{22} \ra{2}= 0,  $&$
  L^{11} | R_2 \rangle + | R_3 \rangle= 0,  $&$
 L^{11} | R_1 \rangle= 0,
$\\&&\\$
  L^{12} | R_1 \rangle - | R_3 \rangle= 0,   $&$
  L^{12} | R_2 \rangle + | R_4 \rangle= 0,   $&$
 L^{22} | R_2 \rangle= 0,
$\\&&\\$
  L^{22} | R_1 \rangle -| R_4 \rangle= 0,  $&$
  T | R_3 \rangle - | R_4 \rangle= 0,  $&$
  T \ra{5}= 0,
$\\&&\\$
  L^{11} \ra{5}= 0,  $&$
  L^{12} \ra{5}= 0,  $&$
  L^{22} \ra{5}= 0,
$\\&&\\$
  L^{11} | R_3 \rangle= 0, $&$
  L^{12} | R_3 \rangle= 0, $&$
  L^{22} | R_3 \rangle= 0,
$\\&&\\$
  T | R_4 \rangle= 0,      $&$
  L^{11} | R_4 \rangle= 0, $&$
  L^{12} | R_4 \rangle= 0,
$\\&&\\
$
 L^{12} \ra{0} + \ra{3} + \ra{4}= 0 , $&$ L^{22} | R_4 \rangle= 0.$&$
$
\end{tabular}}{\phantom{0}}
\end{equation}
Due to this large but nevertheless, consistent system of equations
all auxiliary fields entering in the expansion
of the state vector \p{vector}  are expressed
via the basic field $\ra{0}$
\begin{eqnarray}
 \ra{1} &=& -L^{11} \ra{0}, \quad
\ra{2} = -L^{22}\ra{0},  \quad
\ra{3} = -\frac{1}{2}L^{12}\ra{0},
\nonumber \\
&&\ra{4} = -\frac{1}{2}L^{12}\ra{0},  \quad
\ra{5} = -L^{11}  L^{22}\ra{0},
\nonumber \\
&&|R_4 \rangle =
      (L^{1}  L^{22}   - \frac{1}{2} L^{2}  L^{12}   +
        L^{+1}  L^{11}  L^{22}) \ra{0},  \nonumber \\
&&|R_3 \rangle =
      (\frac{1}{2} L^{1}  L^{12}   - L^{2}  L^{11}   -
        L^{+2}  L^{11}  L^{22})\ra{0},  \nonumber \\
&&|R_2 \rangle=
(L^{2}   + \frac{1}{2} L^{+1}  L^{12}   +
        L^{+2}  L^{22})\ra{0},  \nonumber \\
&&|R_1 \rangle=
      (L^{1}   + L^{+1}  L^{11}   +
        \frac{1}{2}\ L^{+2}  L^{12}) \ra{0}.
\end{eqnarray}

\end{document}